Hindawi

## Review Article
# Importance of Molecular Interactions in Colloidal Dispersions

### R. López-Esparza,[1,2] M. A. Balderas Altamirano,[1] E. Pérez,[1] and A. Gama Goicochea[1,3]

[1]*Instituto de Física, Universidad Autónoma de San Luis Potosí, 78290 San Luis Potosí, SLP, Mexico*
[2]*Departamento de Física, Universidad de Sonora, 83000 Hermosillo, SON, Mexico*
[3]*Innovación y Desarrollo en Materiales Avanzados A. C., Grupo Polynnova, 78211 San Luis Potosí, SLP, Mexico*

Correspondence should be addressed to A. Gama Goicochea; agama@alumni.stanford.edu

We review briefly the concept of colloidal dispersions, their general properties, and some of their most important applications, as well as the basic molecular interactions that give rise to their properties in equilibrium. Similarly, we revisit Brownian motion and hydrodynamic interactions associated with the concept of viscosity of colloidal dispersion. It is argued that the use of modern research tools, such as computer simulations, allows one to predict accurately some macroscopically measurable properties by solving relatively simple models of molecular interactions for a large number of particles. Lastly, as a case study, we report the prediction of rheological properties of polymer brushes using state-of-the-art, coarse-grained computer simulations, which are in excellent agreement with experiments.

## 1. Introduction

British physicist Thomas Graham in 1861 invented the term "colloid" from the Greek word *kolla*, meaning glue. Graham tried to classify the behavior of various chemicals with respect to their diffusion using membranes [1]. Over time, it has become clear that colloids should not be considered as a special class of chemicals but rather correspond to states in which matter is associated, naturally or artificially, due to some physical division. In general, colloidal systems are composed of a continuous medium and a dispersed phase. The case that concerns us is that the colloidal suspension is one in which the continuous medium is in the liquid state (called "solvent") whereas the dispersed phase consists of solid particles (called the "solute"). The size of the solute particles ranges from 10 nm to 1 $\mu$m approximately [2]. Some common examples of colloidal solutions are paints, inks, wastewater, and ceramics [3]. Of course, there are colloidal systems in which the continuous medium is solid or gaseous and the dispersed phase is a liquid or a gas, but these systems are not the focus of this work and therefore shall not be discussed. The present day importance of colloidal suspensions arises from the impact of their industrial applications in contemporary societies. A familiar example of a colloidal suspension, paints, has become an international business that generates billions of dollars annually; therefore there are current basic and applied research efforts devoted to their study and optimization [4]. The process of removing solid particles from waste water to make it drinkable again, that is, removing the solute from the solvent, is a major problem today, intimately related to public health and environmental protection issues. Many industrial processes generate waste in the form of colloidal suspensions, which must be treated and purified before being released to the environment [5]. Moreover, colloidal dispersions are important in academy, and their stability is the result of the delicate balance of attractive and repulsive forces [6], which give rise to complex phenomena of many-body interactions that cannot be fully understood using *ab initio* theories except for some simplified models. Hence, models have been proposed based on effective interactions and computer simulation solutions [7].

## 2. Basic Interactions between Colloidal Particles

The dominant attractive interaction in suspensions of colloidal particles is the van der Waals (vdW) interaction, which is nevertheless short-ranged. It arises from the averaged quantum fluctuations of the electric dipoles on the surface of particles [6]; it is a pair interaction that can be written as

$$U_{\text{vdW}}(r) = -\frac{A}{12\pi} \left(\frac{R}{r}\right)^2,$$ (1)

where $A$ is the so-called Hamaker constant, $R$ is the radius of the colloidal particles (see Figure 1) which are assumed to be identical, and $r$ is the distance between them. The value of $A$ depends on the nature of both the solvent and the colloidal particles [6].

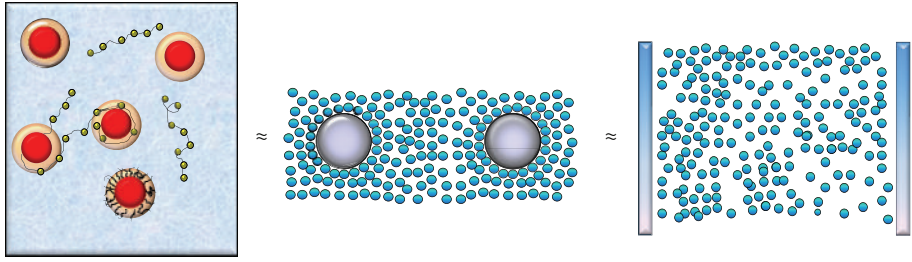

Figure 1: Model of interaction between colloidal particles of radius $R$ and solvent. The leftmost image represents a dispersion of colloidal particles (in red), coated with polymers, which are dispersed with polymer chains (in yellow). The central image illustrates the size difference between the colloids (in gray) and the solvent molecules (in cyan). Because of this size difference, colloidal particles can be considered as planar walls when modeling their interactions with the solvent, as shown in the rightmost image.

Electric charges are usually present on the surface of colloids as well, and such basic interaction must also be accounted for when studying colloidal suspensions. One should recognize that this interaction is expected to be different from the bare Coulomb interaction, because the colloidal particles are dispersed in a liquid medium, which modifies it. A useful expression for the electrostatic interaction in colloidal suspensions, $U_E$, is [6]:

$$U_E(r) = 4\pi\varepsilon \frac{R}{2 + r/a} \psi_0^2 \exp(-\kappa r),$$ (2)

where $\varepsilon$ is the dielectric constant of the liquid medium, $\psi_0$ is the electrical potential on the colloids' surfaces, and $\kappa$ is known as the Debye-Hückel constant. The latter is defined as [6]:

$$\kappa = \left(\frac{e^2 \sum n_{oi} z_i^2}{\varepsilon k_B T}\right)^2,$$ (3)

where $e$ is the fundamental unit of charge; $n_{oi}$ is the concentration and $z_i$ the valence of ions of type $i$, and $k_B T$ is the thermal energy. Equation (3) represents the inverse of the characteristic decay length of the screened electrostatic interaction; see Figure 2, which should be expected to depend on the concentration of electric charges in the medium. Both the vdW and electrostatic interactions are illustrated in Figure 3 by dotted lines.

Since both of these interactions are usually present in most colloidal suspensions, the complex interplay between them has been studied for several decades, giving rise to a compounded potential interaction function, known as the DLVO potential (after Derjaguin, Landau, Verwey and

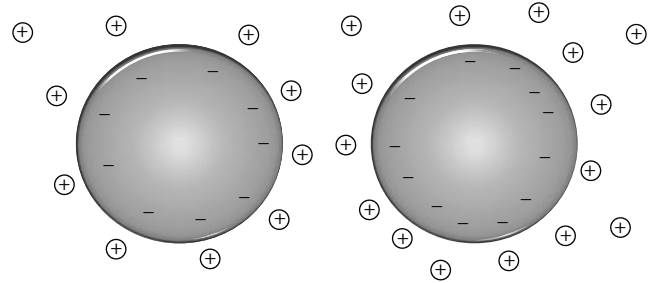

Figure 2: Two colloidal particles (large spheres) in suspension, showing the distribution of ions (small circles) around them, which give rise to screening of the long range electrostatic interaction, yielding (2).

Overbeek), see the solid line in Figure 3. A salient aspect of the DLVO potential is that it develops a repulsive barrier of height $E_{\text{max}}$ at certain relative position $S_{\text{max}}$ between the centers of mass of the colloids. When $E_{\text{max}} \gg k_B T$, those particles whose relative separations are $r > S_{\text{max}}$ are very unlikely to agglomerate, due to Brownian motion for example. The colloidal suspension is, therefore, kinetically stable. When the separation between the particles is relatively short, they are likely to fall into the primary minimum dominated by the vdW attraction (1), leading to the irreversible flocculation of the suspension. There is, of course, a continuous relative-separation distance distribution of the particles at any finite temperature, with some of those distances being large enough to avoid agglomeration. However, the peak of such distribution will be at distances where the energy is at its minimum, and as time progresses more and more particles will fall into that minimum energy state. Given a large enough period of time all particles will move toward their global minimum energy state, leading to the flocculation of the suspension, although the energy barrier $E_{\text{max}}$ may be large enough to prevent such event from happening during a long enough period of time for the suspension to be practically stable. To further increase the kinetic stability period of the suspension, polymer chains are typically grafted on the colloids' surfaces, which introduces an additional repulsive force at short interparticle distances [6].

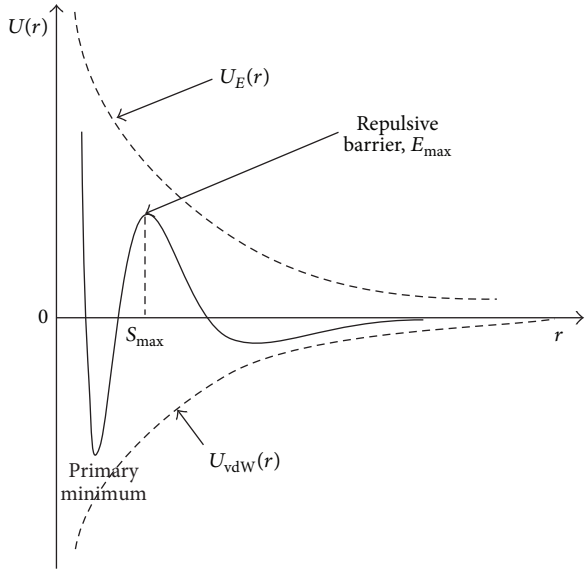

FIGURE 3: The competition between the electrostatic repulsion, $U_E(r)$, and the van der Waals attraction, $U_{vdW}(r)$, leads to the so-called DLVO interaction (solid line), which develops a repulsive barrier of height $E_{max}$ at the position $S_{max}$. Adapted from [3].

When the particles are coated with a polymer layer, which can occur by physical adsorption or when the polymers are anchored by means of chemical bonds, a repulsive force arises at short distance between them [6]. This happens as a result of the tendency of the polymer layer to become more compact, instead of mixing with the neighboring particle. The mechanism that gives rise to this repulsion is entropic and can be understood, for example, by the following argument. Suppose two surfaces from neighboring particles are coated with polymers grafted to the surfaces by only one end, forming "polymer brushes." When two particles coated with polymeric brushes approach, *steric repulsion* arises, which is the result of two factors, schematically shown in Figure 4. The first of them, shown in Figure 4(a), is the saturation of polymers in the region where brushes overlap, which induces an increase in the local osmotic pressure as a result of the disparity in concentrations of polymer and solvent. The other factor takes place when the polymer chains in the region between two particles that are very close to each other cannot take as many possible conformations as when they are far apart, which is known as the "excluded volume effect" (see Figure 4(b)) [7].

This reduction in the degrees of freedom, or loss of entropy, is equivalent to an increase of the free energy of the system, which is not thermodynamically favorable. Therefore, the particles tend to separate to reduce the increase in free energy. The collective result of this mechanism is to produce an effective repulsion between the particles, as illustrated in Figure 5, which shows that factors such as an increase in the density of grafted polymers per unit area per particle or the size of the polymer chains contribute to the stabilization of the colloidal suspension. Figure 5, which shows the interaction between particles coated with polymeric brushes on the $y$-axis as a function of the distance between the particles ($x$-axis), illustrates how repulsion appears when distances between polymeric brushes are short (leftmost curve). By increasing the length of the polymer chains, maintaining the fixed number of chains per unit area on the particles, the repulsion also increases (middle curve). Finally, if the length of the chains is kept fixed but the grafting density of chains per particle increases, the repulsion is even higher (rightmost curve).

## 3. Hydrodynamic and Brownian Interactions in Colloidal Suspensions

So far we have discussed some of the most important interactions in colloidal suspensions in equilibrium, that is, when there are no external forces applied to the system, so that none of the thermodynamic properties are time dependent. When an external force acts on the suspension, the hydrodynamic and Brownian forces are predominant (at finite temperature).

*3.1. Hydrodynamic Forces.* This force appears in a colloidal suspension as a result of the flow of solvent and originates a change in the dynamics of the particles dissolved. Figure 6 illustrates the case where a colloidal particle is subjected to hydrodynamic forces.

The hydrodynamic force can be calculated assuming that (a) the flow is steady, so that the velocity $v_o$ in Figure 6 is constant; (b) the viscous forces dominate over the inertial ones; (c) there are no external forces other than that causing the flow. In that case, the force on the colloidal particle is given by [9, 10]

$$F_H = 6\pi\eta a v_0, \tag{4}$$

where $\eta$ is the intrinsic viscosity of the fluid, considered here as a continuous medium. Equation (4) is also known as "Stokes drag" in honor of George Stokes, who derived it in 1851, and it should be stressed that it is valid only when the colloidal particles are sufficiently far apart so that the hydrodynamic interaction between those particles can be neglected. If the medium is water ($\eta \approx 10^{-3}$ Ns/m$^2$), the hydrodynamic drag force experienced by a particle of one micron that moves at a rate of one micron per second will be $F_H \approx 20$ fN.

*3.2. Brownian Motion.* Brownian motion is characterized by erratic motions driven by the temperature of the fluid and results from collisions between the molecules of the solvent, which is no longer considered a continuous medium. Hence, with increasing temperature the Brownian collisions also increase. The mean energy per particle due to Brownian motion assuming identical and perfectly spherical particles is $E = (3/2)k_B T$. Therefore, the characteristic Brownian force acting on a colloidal particle at a certain temperature can be written (up to a factor of order one) as $F_B \approx E/a$; namely,

$$F_B = \frac{k_B T}{a}. \tag{5}$$

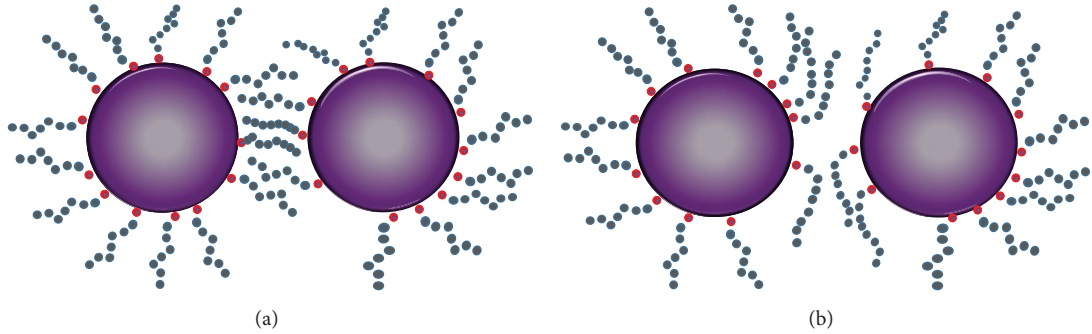

Figure 4: Stabilization of colloidal particles by steric repulsion. (a) Increased local osmotic pressure. (b) Excluded volume effect. Adapted from [3]. Colloidal particles are shown in purple; polymers are the gray chains with red heads.

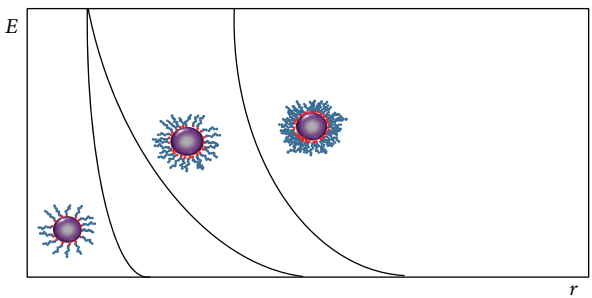

Figure 5: Interaction energy ($E$) as a function of distance ($r$) separating colloidal particles due to the steric hindrance caused by brushes of grafted polymer layers on their surfaces, increasing the length of the polymer, that is, the degree of polymerization and the number of grafted chains per particle. Adapted from [8].

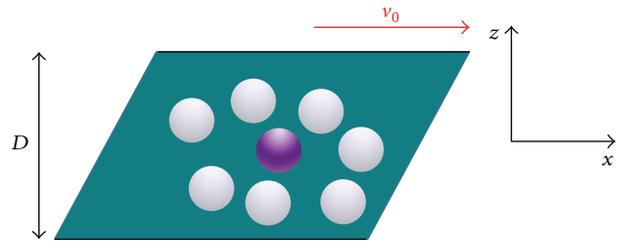

Figure 7: Deformation of a fluid element composed of colloidal particles (gray) suspended in a continuous medium (green). The fluid element is being deformed by an external force, which establishes a steady flow with velocity $v_0$ between two separated surfaces by a distance $D$. The particle's dynamics, in purple, is affected by collisions with its neighboring particles.

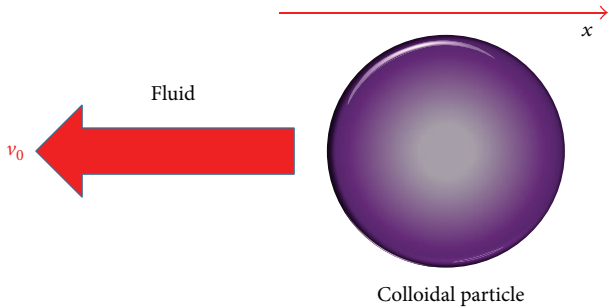

Figure 6: Colloidal particle of radius $a$ immersed in a fluid, which moves in the $x$-direction at a constant speed $v_o$.

The contribution of Brownian motion is more pronounced for particles whose size is of the order of $1\,\mu$m or less; for example, at a temperature of 25°C the magnitude of the Brownian force is about 4 fN; see (5). This type of motion is known as a random walk whose length is given by the number of collisions, $N$, times the diameter of the colloidal particles [8]. Note that up to this point the influence of gravity in a colloidal suspension has not been discussed. Assuming that the particles have a density $\rho \sim 10^2\,\mathrm{kg/m^3}$ and that they are perfectly spherical with a radius equal to $1\,\mu$m, the gravitational force that acts on them is about 4 fN and therefore is comparable to the Brownian strength or may be even less, depending on the density of the colloids. The study of the rheological properties of colloidal suspensions becomes complicated when the particle concentration is high enough for collisions to occur frequently between them, when the particles are not spherical and when additives like polymers are also present. We focus here on a rheological property, namely, the viscosity, which we briefly discuss in what follows.

## 4. Viscosity of Simple fluids

For an operational definition of the viscosity of a fluid, consider an element of it that consists of a cubic volume. This element is subjected to external forces, which deform it into a parallelepiped shape, with the deformation increasing with time. Figure 7 schematically shows this system. In the fluid contained between two parallel surfaces shown in Figure 7 the upper plate moves with constant velocity $v_0$ in the $x$ direction while the one at bottom remains stationary. At speeds low enough to avoid turbulence, the flow will be parallel to the plates. Local velocities in the direction of the flow ($x$) vary linearly along the distance perpendicular to the plates, being zero at the lower plate since it does not move, and $v_0$ at the top plate. Therefore, there is a velocity gradient in the direction perpendicular to the plane of the plates, which is assigned to the $z$-axis; such gradient can be written as

$$\frac{dv_x}{dz} = \frac{v_0}{D} = \dot{\gamma}. \tag{6}$$

In (6) the ratio $v_0/D$ has been defined as $\dot{\gamma}$, which is called the "shear rate" and it is constant if the distance between the plates ($D$) and the speed at which the top plate moves ($v_0$) are constant. To generate flow it is necessary to apply a force to the top plate shown in Figure 7, $F_{xz}$, where the first subscript ($x$) represents the direction of force and the second, ($z$), indicates that the force is applied in the plane normal to the $z$-axis. The force required to move the top plane at $v_0$ velocity is proportional to the area of the plane $A$, so that the dynamics will be defined by the force per unit area, also known as shear stress, $\sigma_{xz}$. Newtonian fluids are then defined as those which obey the following relation [11]:

$$\sigma_{xz} = \eta \frac{dv_x}{dz} = \eta \dot{\gamma}. \tag{7}$$

In the above equation there appears the viscosity, $\eta$, of the complex fluid in question. The type of flow shown in Figure 7 and described by (7) is called "Couette flow" [11]. Those fluids for which $\eta$ does not depend on the shear rate, $\dot{\gamma}$, are defined as "Newtonian fluids," while those for which the viscosity is a function of the shear rate, $\eta = \eta(\dot{\gamma})$, are called "non-Newtonian fluids." When the viscosity decreases with increasing shear rate, it is said that the fluid exhibits *shear thinning*. There are also cases for which the viscosity increases with increasing shear rate; that behavior is called shear thickening or dilatant [12].

Some familiar fluids such as blood, yogurt, or paints exhibit pseudoplastic behavior. The viscosity of paints, for example, decreases when the shear rate is increased, so that it costs less physical effort to apply it to a surface when the speed of the brush or roller is increased. On the other hand, when sand is completely wet, it behaves as a dilatant fluid or shear thickener: the higher the shear rate at which it flows, the higher its viscosity. There are other important types of flows in addition to Couette flow, but those are beyond the scope of this work. Additionally, there are also viscoelastic types of flow, when the applied force is dependent upon frequency so that the fluid responds to such force in a highly complex manner [8]; those topics shall be left out of the present work also.

## 5. Modeling the Interaction between Polymers and Surfaces

Finally, we present briefly some recent trends on somewhat more realistic models of suspensions with soft particles. These efforts come mainly from molecular simulations, since the purely analytical approaches have proved to be extremely difficult to solve for many-particle systems.

*5.1. Dissipative Particle Dynamics Simulations.* Among the most successful computer simulation tools used presently it is the technique known as dissipative particle dynamics

(DPD) [13]. The main distinction between DPD and microscopic molecular dynamics simulations [14] is that the force acting between any two particles $i$ and $j$ in DPD is given not only by a conservative force ($\mathbf{F}_{ij}^C$), but also by dissipative ($\mathbf{F}_{ij}^D$) and random ($\mathbf{F}_{ij}^R$) forces. The total force acting on any given pair of particles is the sum of these three forces:

$$\mathbf{F}_{ij} = \sum_{i \neq j}^{N} \left[ \mathbf{F}_{ij}^C + \mathbf{F}_{ij}^D + \mathbf{F}_{ij}^R \right]. \tag{8}$$

The conservative force is given by a soft, linearly decaying function and it is repulsive:

$$\mathbf{F}_{ij}^C = \begin{cases} a_{ij} \left( 1 - r_{ij} \right) \hat{\mathbf{r}}_{ij} & r_{ij} \leq r_c, \\ 0 & r_{ij} > r_c, \end{cases} \tag{9}$$

where $\mathbf{r}_{ij} = \mathbf{r}_i - \mathbf{r}_j$, $r_{ij} = |\mathbf{r}_{ij}|$, $\hat{\mathbf{r}}_{ij} = \mathbf{r}_{ij}/r_{ij}$, $r_{ij}$ is the magnitude of the relative position between particles $i$ and $j$, and $a_{ij}$ is the intensity of the repulsion between any pair of particles. The dissipative and the random forces are, respectively,

$$\mathbf{F}_{ij}^D = -\gamma \omega^D \left( r_{ij} \right) \left[ \hat{\mathbf{r}}_{ij} \cdot \mathbf{v}_{ij} \right] \hat{\mathbf{r}}_{ij} \tag{10}$$

$$\mathbf{F}_{ij}^R = \sigma \omega^R \left( r_{ij} \right) \xi_{ij} \hat{\mathbf{r}}_{ij}, \tag{11}$$

where $\sigma$ is the amplitude of the random force and $\gamma$ is the coefficient of the viscous force. They are related in a way as follows: $k_B T = \sigma^2/2\gamma$, as a manifestation of the fluctuation, dissipation theorem [15]; $k_B$ is Boltzmann's constant and $T$ the absolute temperature, $\mathbf{v}_{ij} = \mathbf{v}_i - \mathbf{v}_j$ is the relative velocity between the particles, and $\xi_{ij} = \xi_{ji}$ is a random number uniformly distributed between 0 and 1 with unit variance. The weight functions $\omega^D$ and $\omega^R$ vanish for $r > r_c$ and are chosen for computational convenience to be (see [16])

$$\omega^D \left( r_{ij} \right) = \left[ \omega^R \left( r_{ij} \right) \right]^2 = \max \left\{ \left( 1 - \frac{r_{ij}}{r_c} \right)^2, 0 \right\}. \tag{12}$$

All forces between particles $i$ and $j$ vanish further than a finite cutoff radius $r_c$, which represents the characteristic length scale of the DPD model and it is usually chosen as $r_c = 1$. We chose $\sigma = 3$ and $\gamma = 4.5$ so that $k_B T = 1$ [17].

The DPD model can be easily adapted to study fluids under confinement, by simply placing surfaces at the ends of the simulation box, say, perpendicularly to the $z$-axis, that is, at $z = 0$ and $z = L_z$. Various models for such effective surfaces have been used [18, 19], but for most purposes it is sufficient to use a simple, linearly decaying force so that it is similar to the particle-particle force law shown in (9); namely,

$$\mathbf{F}_{wall} \left( z_i \right) = \begin{cases} a_w \left( 1 - \frac{z_i}{z_C} \right) \hat{\mathbf{z}} & z_i \leq z_c \\ 0 & z_i > z_c. \end{cases} \tag{13}$$

The value of the cutoff distance $z_C$ is chosen as 1. The role of the wall force in (13) is to act as a geometric constraint for polymer chains and the solvent molecules. Polymer molecules can be modeled as linear chains made up of identical beads which are joined by freely rotating harmonic springs:

$$\mathbf{F}_{\text{spring}}\left(\mathbf{r}_{ij}\right) = -\kappa_0 \left(r_{ij} - r_0\right)\widehat{\mathbf{r}}_{ij}, \tag{14}$$

where the spring constant is $\kappa_0$ and the position of equilibrium is $r_0$. As illustrated in Figure 8, among the advantages of DPD is that it does not solve the motion of particles at the atomic level, but rather at the mesoscopic level [20–23].

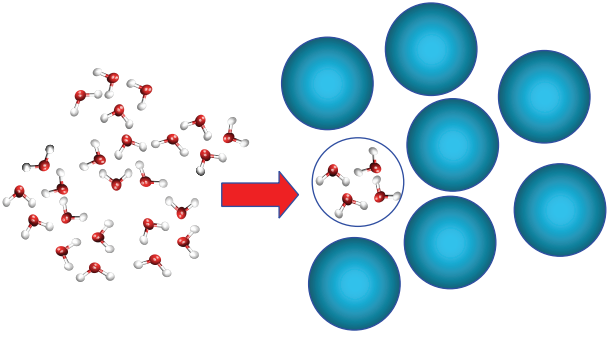

Figure 8: Illustration of the coarse-graining in DPD equal to four water molecules per DPD particle.

We have improved upon earlier models, especially for the case of nonequilibrium simulations [24], as shall be exemplified in the following section.

### 5.2. Application to Polymer Brushes as Friction Reducing Agents.
Some experiments carried out on polymers attached by one of their ends to a surface, thereby forming "polymer brushes," have shown that the friction coefficient between surfaces covered with polymer brushes is considerably smaller than the value obtained for bare surfaces [25]. This result has important potential applications, for example, in the design of lubricants, as slip and antiblocking agents in plastics, and in the pharmaceutical industry when designing and optimizing drug-carrying liposomes, to mention a few. However, the mechanisms that promote such reduction in the friction coefficient when polymer brushes are added are not yet understood, which make of these systems attractive research subjects using computational simulations, particularly with DPD, given the scales of the experiments mentioned above. In what follows we report nonequilibrium DPD simulations of polymer brushes grafted on soft surfaces under steady (Couette) flow. The reason for modeling soft surfaces is that we would like our results to be applicable to cells and liposome covered with brushes under flow. Our model for polymer brushes is illustrated in the left image shown in Figure 9, while the soft surface model is shown in the right image in Figure 9. One end of the polymer is bound to the surface via a covalent bond (dashed circles). The constant force associated with each polymer is modeled with a harmonic force of spring constant $k$. The solvent molecules are shown in red.

The simulation box is made of two parallel substrates placed at the ends of the simulation box on the $xy$-plane, with opposing polymer brushes, separated by a fixed distance of 45 Å. The "heads" of the brushes in the upper layer, shown as dashed circles in the left image in Figure 9, move from left to right at constant velocity $v_0^* = 1.0$, while those at the bottom surface move from right to left with velocity $v_0^* = -1.0$. This produces Couette flow. The area covered by the brushes is $A = 2045$ Å$^2$. We modeled the brushes as linear polymers made up of 7 beads joined by harmonic springs with $\kappa_0 = 100.0$ and $r_0 = 0.7$; see (14). The conservative DPD interaction $a_{ij}$ (see (9)) is set at 78.0 (in reduced DPD units) for particles of the same type: solvent-solvent and polymer-polymer. For solvent-polymer interactions, $a_{ij} = 80.0$; this has been shown to be a good model for polyethylene glycol in water [17]. To promote the adsorption of the polymers' head to the surfaces the constant $a_w$ in (13) was set at 60.0, while for the rest of the polymer monomers and for the solvent $a_w = 120.0$, all in reduced DPD units. DPD simulations in the canonical ensemble (at constant density $\rho^* = 3.0$ and temperature $k_B T^* = 1.0$) were run for $10^2$ blocks of $2 \times 10^4$ time steps ($\delta t^* = 0.01$) each, amounting for a total simulation time of $0.02\ \mu$s. Half of that time was used to equilibrate the system and the second half was used for the production phase.

Keeping the separation between the opposing brushes fixed, as well as the brushes polymerization degree ($N = 7$ beads), we then proceeded to perform simulations at increasing values of the polymer grafting density, $\Gamma$, which is defined as the number of polymer chains per substrate unit area. $\Gamma$ is a variable that can be controlled experimentally; that is why we have chosen it. We calculated two properties, namely, the viscosity ($\eta$) of the fluid, defined by (7), and the coefficient of friction (COF), defined as $\mu = F_x/F_z$, with $F_x$ being the force acting on the beads grafted to the surfaces (dashed circles in Figure 9) along the $x$-direction. The force in the direction perpendicular to the surface is $F_z$ [21]. The results are shown in Figure 10.

The COF shows a strikingly different behavior as a function of $\Gamma$ from that displayed by the viscosity, as seen in Figure 10. The latter increases monotonically with $\Gamma$, while the COF is reduced by almost an order of magnitude with increasing $\Gamma$, reaching its minimum at about $\Gamma = 1.7$ polymer chains/nm$^2$, after which the COF starts growing incipiently. The minimum value of the COF is in excellent agreement with that found by Klein and coworkers on polymer brushes [25]. This model of polymer brush appears to be a very good slip agent for grafting densities smaller than about 2 chains per squared nanometer, although the viscosity of the fluid always increases. The origin of this behavior rests on the fact that the viscosity is proportional to the force acting perpendicularly to the surfaces ($F_z$, see also (7)); this force (per substrate unit area) is the osmotic pressure of the confined fluid. When $\Gamma$ is increased, the osmotic pressure must also increase, since there are an increasing number of polymers in the volume occupied by the solvent. For the monotonic increase in the viscosity with $\Gamma$, see the filled squares in Figure 10. By contrast, the COF is defined as the ratio of the force along the direction of flow ($F_x$) over the force along the direction perpendicular to it ($F_z$); therefore, for moderate grafting densities the osmotic pressure dominates over the force on the beads moving along the flow, and the COF is reduced. However, as $\Gamma$ is increased further, the number of solvent molecules must be reduced because the number of polymers increases (to keep the total density constant). This means that the layer of solvent molecules that was sandwiched between the opposing brushes, which eases the slip of one brush over another, is also reduced. The consequence of this thinning of the solvent layer is an increase in the force of one brush acting on another ($F_x$), which translates as an increase in the COF at large $\Gamma$ (see Figure 10). The advantage of performing DPD simulations is evident from the discussion above, since one has total control over different variables and physicochemical conditions so that an optimized friction reducing agent can be designed and tested, using properly adapted algorithms for many-particle systems.

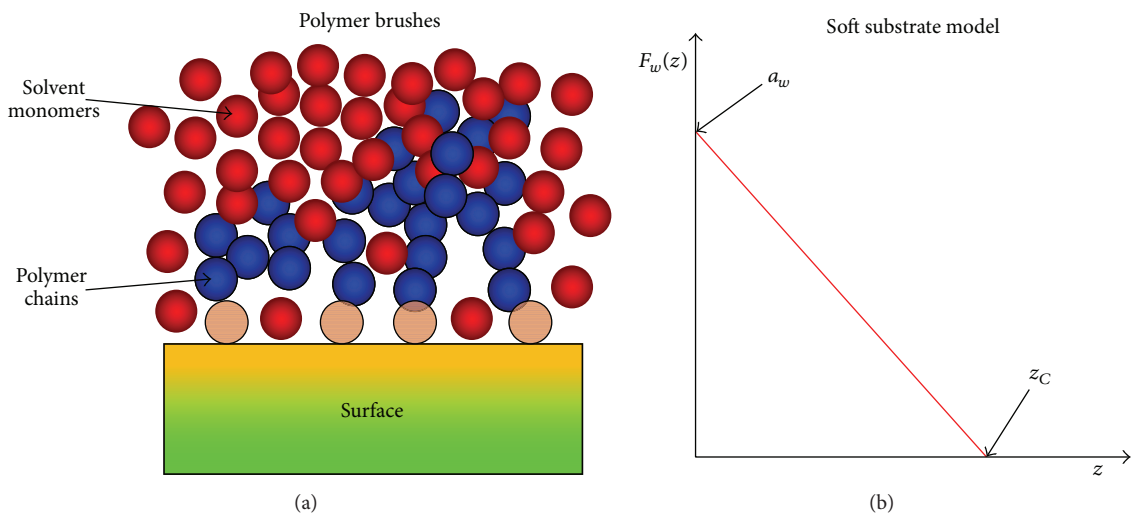

Figure 9: (a) Model for polymer brushes grafted on a substrate. (b) A simple model for the soft surface effective interaction with the rest of the fluid; $a_w$ is the magnitude of the interaction and $z_C$ is the cutoff distance, beyond which the surface interaction vanishes; see (13).

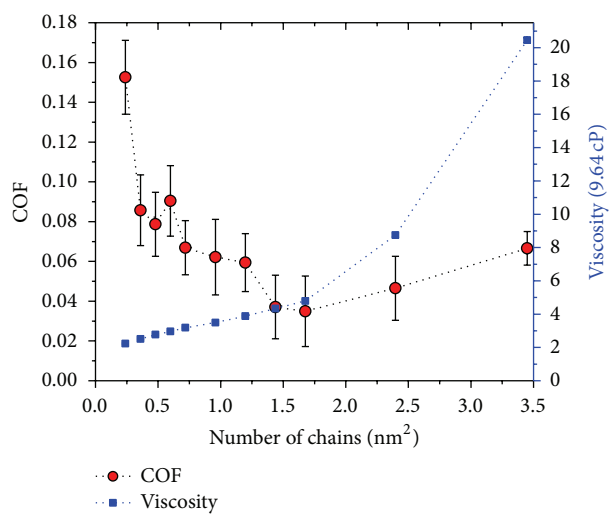

Figure 10: The coefficient of friction (COF, filled circles) and the viscosity (filled squares) of a fluid made up of two opposing polymer brushes with degree of polymerization equal to 7 beads, immersed in a solvent, as functions of the grafting density of polymer chains. The error bars for the viscosity are smaller than the symbols' size. Dotted lines are only guides for the eye.

## 6. Conclusions

We have reviewed the basic interactions and properties of colloid dispersions in equilibrium, as well as hydrodynamic and Brownian forces. Emphasis was placed particularly on the case of interactions between polymers and surfaces, which are systems of widespread current interest in the scientific community. We focused also on the application of state-of-the-art, coarse-grained computer simulations as tools to study those interactions in complex soft matter systems. This is illustrated with an application to the design and test of optimal friction reducing agents, whose rheological properties were predicted. This work should be useful for graduate students in physics, materials science, chemistry, and chemical engineering wishing to get firsthand experience with molecular interactions and their macroscopic consequences in complex fluids using relatively simple models and modern research tools to solve them, such as computer simulations.

## Conflict of Interests

The authors declare that there is no conflict of interests regarding the publication of this paper.

## Acknowledgments

A. Gama Goicochea acknowledges C. Ávila, J. D. Hernández, C. Pastorino, Z. Quiñones, E. Rivera, and M. A. Waldo for instructional input. A. Gama Goicochea, M. A. Balderas Altamirano, and R. López-Esparza thank IFUASLP for its hospitality and Universidad de Sonora (UNISON) for permission to run simulations in their *Ocotillo* cluster. R. López-Esparza thanks UNISON for sabbatical support.

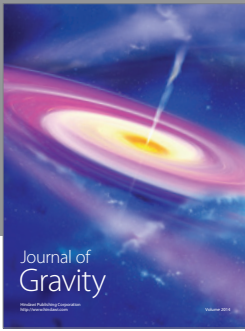

Journal of
Gravity

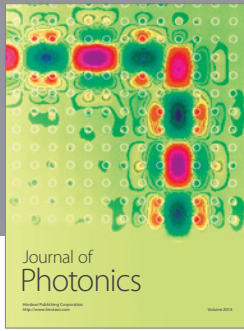

Journal of
Photonics

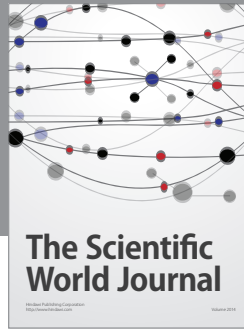

The Scientific
World Journal

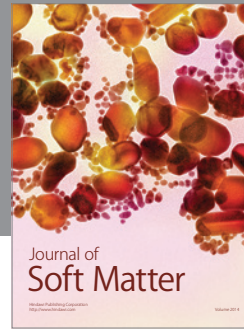

Journal of
Soft Matter

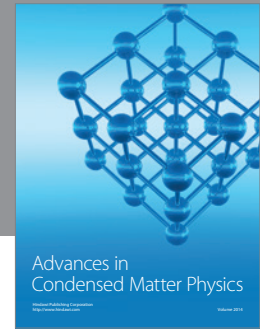

Advances in
Condensed Matter Physics

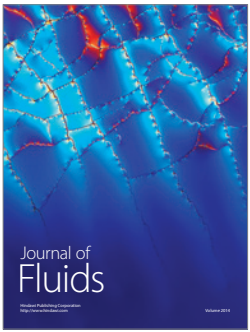

Journal of
Fluids

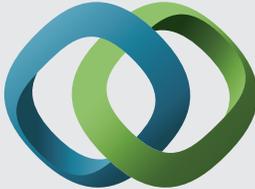

Hindawi

Submit your manuscripts at
http://www.hindawi.com

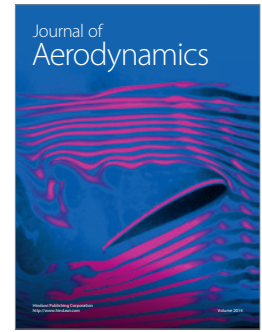

Journal of
Aerodynamics

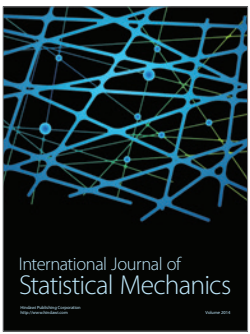

International Journal of
Statistical Mechanics

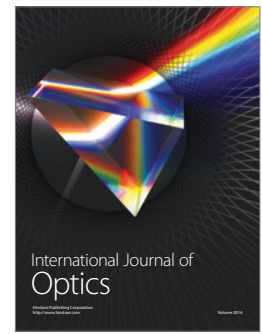

International Journal of
Optics

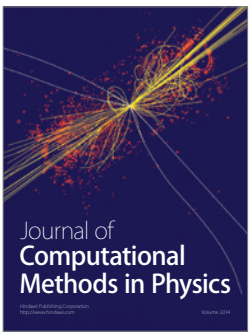

Journal of
Computational
Methods in Physics

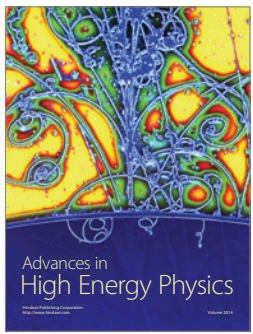

Advances in
High Energy Physics

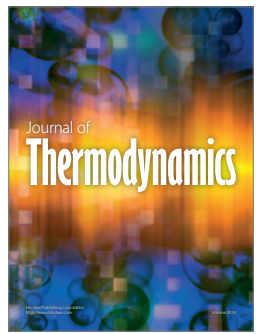

Journal of
Thermodynamics

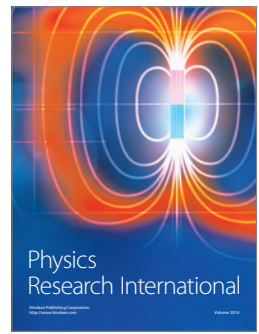

Physics
Research International

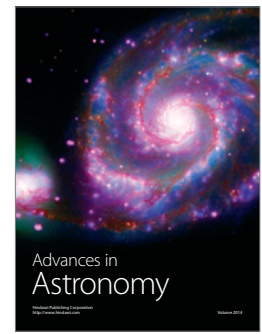

Advances in
Astronomy

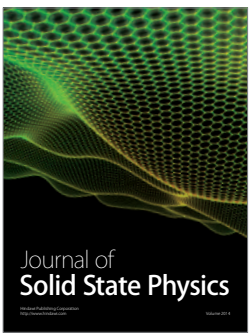

Journal of
Solid State Physics

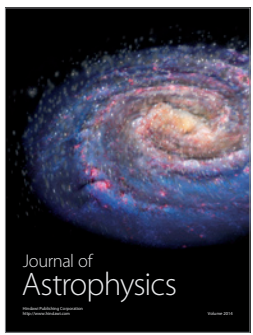

Journal of
Astrophysics

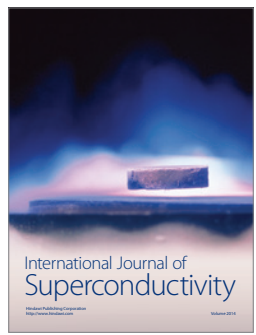

International Journal of
Superconductivity

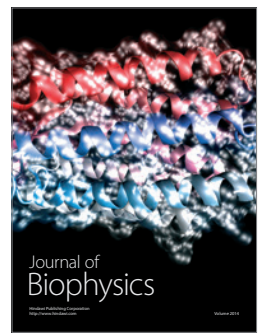

Journal of
Biophysics

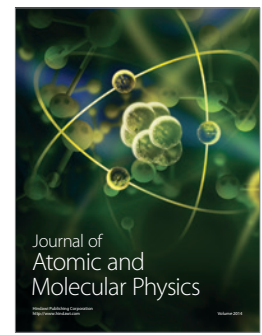

Journal of
Atomic and
Molecular Physics